\documentstyle[epsfig,12pt]{article}
\begin{document}

\begin{center}
{\Large\bf Periodicities in the occurrence of solar coronal mass
ejections }
\end{center}
\begin{center}
{\Large\bf K. M. Hiremath}
\end{center}
\begin{center}
{\em Indian Institute of Astrophysics, Bangalore-560034, India,
E-mail : hiremath@iiap.res.in}
\end{center}

\begin{abstract}
Recent overwhelming evidences (Hiremath 2009a and references there in) show
that sun indeed strongly influences the earth's climate and environment.
In addition to well known sunspot activity, right from the era of discovery
by Galileo, other activity phenomena such as flares and coronal mass
ejections are also the main drivers that input vast amount of energy, mass
and momentum into the earth's environment.
Solar coronal mass ejections (CME) are dynamic events of blob of plasma which
originate from the sun and affect the natural environment of the earth and the
planets. These events involve significant ejection of masses from the sun, typically
$10^{15}$ to $10^{16}$ grams and energies of the $\sim$ $10^{31}$ to $10^{32}$ ergs. These disturbing events
play a major role in producing storms in the earth's magnetosphere and
ionosphere, which in turn are responsible for the enhanced auroral activity,
satellite damage and power station failures on the earth. It would be useful to
minimize such disasters in case CME occurrences found to be periodic in nature
and detected well in advance. The Fourier analysis of the CME
occurrence data observed by the SOHO satellite shows significant power around
1.9 yr., 1.2 yr., 265 day, 39 day and 26 day periodicities which are almost similar to
the periodicities detected in the Fourier analysis of underlying activities of the
photosphere.  The wavelet analysis of CME occurrences also shows significant
power around such periods which occur near the peak of solar activity. For the
sake of comparison, the occurrences of H-alpha flares are subjected to Fourier and
wavelet analyses.  The well-known periods (1.3 yr., 152 day, 27 day) in the flare
occurrences are detected. The wavelet analyses of both the occurrences yield
the following results : (i) in the both CME and Flare activities,
long period ( $\sim$ 1.3 yr) activity occurs around solar maximum only and,
(ii) flare activity of long period, especially for the period 1.3 year, lags behind the
long period CME activity nearly by six months.
Possible physical explanation for the 1.2 yr CME quasi-periodicity
is briefly discussed.
\end{abstract}

\noindent Keywords:  {\em Sun: corona-Sun: coronal mass ejections
(CMEs)-Sun: flares-Sun: periodicities}

\section{Introduction}
    On the sun, periodic phenomena have been observed with a wide variety of
periods ranging from minutes to decades and perhaps even centuries (Hiremath
1994 and references there in). A well-known
periodicity around 5 minutes has been detected from the analysis of the velocity
dopplergrams (Leighton 1960)  which are coherent oscillations on the sun. 
Other conspicuous periodicity of near 27 day, which may be due to solar 
equatorial rotation period, has been found in the several activity phenomena. This 
periodicity is noticed 
in the sunspot number (Lean 1991; Balthasar 2007; Kilic 2009), in the sunspot areas (Kilic 2009),
in the HeI-1084 nm equivalent width, 
the plage index and the 10.7 cm radio flux. The near 27 day periodicity is also detected
in the following activity phenomena : Ca K plage area (Singh and Prabhu 1985), globally 
averaged coronal radio fluxes (Kane, Vats and Sawant 2001; Kane 2002; Hiremath 2002)
and in the CME activity during Solar Cycle 23 (Lara {\em et. al.} 2008). 

The near 150 day periodicity has been detected in the solar flare occurrences (Rieger 1984; Ichimota 1985; Bai 1987; Droge 1990), in the 
sunspot areas (Lean 1990; Carbonell and Ballester 1990; Oliver, Ballester and Baudin 
1998; Krivova and Solanki 2002; Chowdhury, Manoranjan and Ray 2009) , 
in the Zurich sunspot number (Lean and Brueckner 
1989), in the radio bursts of type II and IV (Verma {\em et. al.} 1991),
in the solar neutrino fluxes (Sturrock, Walther and Wheatland 1997), in
the solar diameter (Delache {\em et. al.} 1985; Ribes {\em et. al.} 1989),
in the x-ray flares (Dimitropoulou, Moussas and Strintzi 2008), 
in the surface mean rotation (Javaraiah and Komm 1999; Javaraiah 2000)
 and in the photospheric magnetic flux (Ballester, Oliver and Carbonell 2002).  From the solar activity indices, the long periods in the range of 240-330 days also 
have been detected. Recently, Hiremath (2002) Fourier analyzed the unequally 
spaced data of globally averaged radio fluxes (at 275, 405, 670, 810, 925,
1080, 1215, 1350, 1620 and 1755 MHz) observed by the Cracova Observatory and
detected a 274 day periodicity for all the observed radio fluxes. 
Javaraiah and Komm (1999) detected the 243 day periodicity in the photospheric mean rotation determined from the Mount Wilson Velocity data.  By studying the monthly 
mean of the Zurich sunspot number from 1749 to 1979, Wolff (1983) 
reported a peak in the power spectrum 
at 323 days. Delache {\em et. al.} (1985)  found this peak in the power spectrum of the solar 
diameter measurements from 1975 to 1984. Pap et. al (1990)., detected periodicities 
between 240-330 days in the 10.7 cm radio flux, the CaK plage index, the UV flux at 
Lyman-alpha and MgII core to the wing ratio. 

From the helioseismic data, the 1.3 yr.  
periodicity is detected near base of the
convection zone (Howe, {\em et. al.} 2000). However, using same helioseismic data, 
Antia and Basu (2000) conclude that there is no 1.3 yr. periodicity near the base 
of the convection zone. The near 1.3 yr
periodicity is also detected in the sunspot data (Krivova and Solanki 2002), 
in the photospheric mean rotation (Javaraiah and Komm 1999; Javaraiah 2000),
in the magnetic fields inferred from H-alpha filaments (Obridko and Shelting 2007),
in the large-scale photospheric magnetic fields (Knaack, Stenflo and Berdyugina 2005)
and in the green coronal emission line (Vecchio and Carbone 2009). The 
periodicities in the range of 500-550 days 
have been claimed in the different manifestations of the 
solar activity (Oliver, Carbonell and Ballester 1992).
 Going through all the afore mentioned studies and by finding the periodicities in
the solar activity indices, one may also expect such periodicities in the CME
occurrences. Infact such a periodic analysis (Lou {\em et. al.} 2003) 
of CME occurrences yields the periods around 358, 272 and 196 days. 
Aim of the present study is two fold: (i) to search for the
periodicities in the occurrence data and, (ii) if such periods are detected, whether they
are continuous through out the time of observation or not. Moreover, study of finding
the periodicities of CME occurrences will be useful in order to minimize the following
disasters: in the space navigation, reduction in the life of a satellite and power failures in the electric grids on the earth, etc.  

\section{ Data and Analysis }

For the present analysis, we use the CME occurrence data ( from the years 1996 to
2001) observed by the SOHO
satellite from the on board coronal LASCO experiment. The same data has been
compiled by Gopalswamy, Yashiro and Michalek ( $http://cdaw.gsfc.nasa.gov/CME-List$). 
There are three months data gap during July 
1998-Sept 1998 and one month data gap in Jan 1999. For the
sake of comparison of the periodicities and filling the data gap in CME occurrences, we
use the H-alpha flare activity occurrences which is closely associated with the CME
activity (Harrison 1995).  The H-alpha flare occurrence data is available at
$http://ftp.ngdc.noaa.gov/STP/SLAR-DATA/SOLAR-FLARES/HALPHA-FLARES$.
 In order to have reliable statistics of the occurrences of the CME, 
we rebin (collect) the data in 10- day intervals. 
 In the beginning of Jan and Feb 1996, CME occurrences
are rare even during 10- day intervals. Hence, we consider the data
around middle (12th ) of March 1996 onwards. Excluding the gaped data of four
months, totally we have 1960 day of observations
and for the 10- day intervals, we have 196 data points for the following
analysis. 

In Fig 1, we present the occurrences of CME (in blue color) 
and flares (in red color) normalized to their respective means. After computing 
deviation from the mean of the CME occurrences, we fit a sine curve of the form 
$Asin(\omega t + \phi)$.  Here A is the amplitude, $\omega\, (= 2 \pi /T,\, T\, is\, 
the\, period)$ is the frequency and $\phi$ is phase angle of the sine wave. Fitting a
sine curve to the deviation data is a process of removing a long 
term trend of 11 years, although fitting of sine curve is not right
as the profile of solar cycle is similar to profile
of a solution obtained from the forced and damped harmonic
oscillator (Hiremath 2006).
In Fig 2, we present the deviation from the mean of the
CME occurrences with a superposition of fitted sine curve (continuous line with a red
color). We then compute the difference from the fitted sine curve and deviation from the
mean data and the result is presented in Fig 3. 

Owing to
gaps in the data of the CME occurrences, we apply the Scargle (1989) method of Fourier
analysis for the unevenly spaced data and whose Fourier transform is presented in 
Fig 4. We detect the periodicities 1.45 yr., 353 days, 235 days, 38 days
and 26 days with the powers whose significance are greater than 3 $\sigma$ level. Though
periods with significant powers appear to be almost similar to the periods detected in
other solar activity indices, a large gap of 90 days and 30 days may act as sine waves
and may induce spurious periods in the Fourier analysis. For example, spurious periods
may be detected in the multiples (e.g. for the 90 day gaps, we may get 270 day and 360
day) of harmonics. In order to rule out such spurious periods, one way is to
interpolate between the data gaps and make the data of equal intervals for applying
the usual Fast Fourier transform (FFT). Since the duration of the data gap is very 
large, interpolation method may not fill accurately the data gaps. Another way
to solve this problem is as follows. It is well established that solar x-ray 
flares are well associated 
with the coronal mass ejections (Harrison 1995). Assuming that such association may also 
exists (in the following analysis we find a very good correlation)  between the H- alpha flare and the CME occurrence, we use H-alpha flare 
occurrence data in the following analysis for the correlative study. Then from
the linear least square fit between the CME and H-alpha flare occurrences,
we fill the data gaps for applying the usual FFT for the subsequent analysis. 

In Fig 5, we present
the relation between the occurrence of CME and flare data normalized to their
respective means. Notice a very good correlation ( 65 \% ) of CME occurrences with the
H-alpha flare occurrences. From the $z$ statistics (Fisher 1930), the probability
of the correlation coefficient is found to be significant below less
than 1 $\%$ indicating a very good correlation. Presently, we can not conclude 
whether production of CMEs lead
to the production of the H-alpha flares and {\em vice. verse}. It appears that
the occurrences of CME and H-alpha flare activities have one to one correspondence
and may be symptom of the same physical phenomenon in different activities (
see also Zang {\em et. al.,} 2001). 
 The occurrences of CME and flare data are fitted with a
straight line (red continuous line superposed on  
Fig 5) of the form CME = A + B (flare), where 'CME' is the CME occurrences 
and 'flare' is the H-alpha flare occurrences, A = $0.56 \pm 0.01$ and 
B = $0.44 \pm 0.08$. Using this
law and known flare occurrences during the months of July 1998-Sept 1998 and Jan
1999, we fill the data gaps of the CME occurrences during those months. Totally
we have 2120 day of observations and 212 data points for the 10 -day
intervals. Then we fit a
sine curve to the deviation from the mean of the filled CME occurrence data. The
difference between deviation from the mean and the fit of the sine curve is presented in
 Fig 6. Except for the period of July 1998-Sept 1998 and Jan 1999,
this plot is almost similar to the plot presented in the Figure 3.

In Fig 7, we present Fourier power spectrum of the CME occurrences for the
equally spaced data of the resulting difference from the sine wave fit. From this
analysis, we recovered all the periodicities as in the power spectrum of the gaped data
with a much better peak at 1.2 yr.  periodicity which is detected in almost all of the solar
activity indices. It is surprising that the periodicity around 150 days, found in most of
the activity indices, is absent in the CME occurrences even with the filled data. In fact
we Fourier analyzed separately the difference data for the years March 1996-June 1998 and
Feb 1999-Dec 2001 and, we could not detect the periodicity around 150 days. This
absence of around 150 days periodicity is also consistent with the previous
study (Lou {\em et. al.} 2003). In 
 Fig 8, we present Fourier power spectrum of the flare data and we
detect the periodicities 1.3yr. 212day, 152 day, 133 day, 82 day, 62 day and 27 day
whose powers are greater than 3 $\sigma$ confidence level.

In order to know whether the detected periods of either CME or flare occurrences
are continuous through out the observed period, we subject both the difference data set
of filled CME and flare occurrences to the wavelet analysis. In Fig 9,
we present power spectrum of the wavelet analysis of the filled CME occurrences and in
 Fig 10, we present the wavelet power spectrum of the occurrences of
the flares. The decreasing order of wavelet power is represented as follows : the highest
 power in red, the next power in pink, the green and the blue having the least powers. 
The wavelet power in the gray hatched is not significant due to the fact that
the data has been padded with zeros before applying the wavelet spectrum. 
It is crucial to be noted that all the periodicities detected from the Fourier
analyses are also present in the wavelet power spectrum (concentrations of the wavelet
power in the interior regions of light blue contours, which are significant at 90 \%
confidence levels). One disadvantage of the wavelet power spectrum is that we can
not separate the periodic structures as in usual FFT method. 
It is to be noted that the activity of the CME and flare occurrences
for the short periods, below one year, start simultaneously near the maximum of the
solar activity. However, the activity of the long period H-alpha flare activity, especially
1.3 yr., begin around the maximum period has a phase lag of nearly six months 
compared to the long periods of CME activity. {\em To be precise, the flare 
activity of long periods lags behind the CME activity of the long periods nearly 
by six months.} Another interesting result to be noted is that all the dominant
near 150 day and 1.3 year periodicities of the flare activity excite simultaneously
around the middle of maximum year 2000. On the other hand, the CME activity of 
the long periods ( 1.2 and 1.9 year ) excite in different times. 

\section {Conclusions and Discussion}
From the constraint of the CME data set from 1996-2001, presently it is not clear
whether CME periodic activity of short and long periods behaved similarly in the
previous solar cycles. Analyses of either future observed cycles of CME activity or
analyses of proxy data such as flares, microwave bursts and auroras may give clear
picture of the CME occurrence periodic activity. Over all conclusion from the present
analysis is that the FFT and the wavelet analyses of the CME occurrences lead to
discovery of the periodicities 1.92 yr., 1.2 yr., 265 day, 39 day and 26 day which are
almost similar to the well known periodicities found in different solar activity 
indices. It is known fact (Olive, J.P,
Chaloupy, M \& Schweitzer 2002) that the station
keeping maneuvers are limited to 4-6 months and SOHO orbit has
6-month period. In fact we have a peak (from the frequency
range 0.035-0.052 in Fig 4 \& from the frequency range 0.0083-0.0056
in Fig 7) around
4-6 months which is below the detection ( 3$\sigma$) level. From the wavelet analysis, it is found that the long period CME 
activity starts near the solar maximum. The comparative study of the wavelet power 
spectrums of the occurrences of the CME and the flare data
suggests that the long period flare activity lags behind the long period CME activity
nearly by six months. 

As for the genesis of the long periodicities
detected in the CME occurrence, some more years of occurrence data is required
to give any meaningful physical interpretation. However, the 1.2 yr quasi-periodicity
is almost similar to 1.3 yr quasi-periodicity that is discovered in
many indices of solar cycle and activity phenomena. Recently Hiremath (2009b)
has proposed new ideas on the physics of the solar cycle. It is
proposed that sun is pervaded by a combined weak poioidal and strong toroidal 
magnetic field structure of primordial origin whose diffusion time scales
are $\sim$ billion yrs. While explaining the near 11 yr sunspot
 (or 22 yr magnetic) cycle, physical inferences suggest that
near 1.3 yr periodicity is due to long period Alfven wave perturbations
to the ambient strong toroidal magnetic field structure near
base of convection zone that travel to the surface near maximum
of solar cycle and occur near the 20-25 deg latitude (or 70-75 deg
 colatitude) zone. Further details can be found from
that study (see section 5.2.2 of Hiremath 2009b).

\begin{center}
{\Large\bf Acknowledgments}
\end{center}
This paper is dedicated to my beloved parents who constantly encouraged my
research carrier when they were alive.
The CME catalog  is generated and maintained by the center for Solar physics and
Space Weather, The Catholic University of America in Cooperation with the Naval Research Laboratory
and NASA. SOHO is a project of international cooperation between ESA and NASA. We used the
wavelet software developed by C. Torrence and G. Compo  and is available at the web site
http://paos.colorado.edu/research/wavelets/.

\section {\Large\bf References}

\noindent Antia, H. M. \& Basu, S., 2000, ApJ, 541, 442

\noindent Bai, T, 1987, ApJ, 318, L85

\noindent Ballester, J. L., \& Baudin, F., 1998, Nature, 394, 552

\noindent Ballester, J. L., Oliver, R. \& Carbonell, M., 2002, ApJ, 566, 505

\noindent Balthasar, H., 2007, A\&A, 471, 281

\noindent Carbonell, M. \& Ballester, J. L., 1990,  A \& A, 238, 377

\noindent Chowdhury, P., Manoranjan, K and Ray, P. C., 2009, MNRAS, 392, 1159

\noindent Delache, P. Laclare, F and Sadsaoud, H., 1985, Nature, 317, 416	

\noindent Dimitropoulou, M., Moussas, X and Strintzi, D., 2008, MNRAS, 386, 2278

\noindent Droge, W. et. al., 1990, Astrophys. J, Supp, 73, 279

\noindent Fisher, R. A., 1930, in Statistical Methods for Research Workers, p. 74 \& p. 168

\noindent Harrison, R. A., 1995, A\& A, 304, 585

\noindent Hiremath, K. M., 1994, Ph. D Thesis, Study of Sun's long period oscillations, Bangalore University, India

\noindent Hiremath, K. M., 2002, in Proc. IAU Coll 188, (ESA SP-505), p. 425

\noindent Hiremath, K. M., 2006, A\&A, 452, 591

\noindent Hiremath, K. M., 2009a, accepted in Sun and Geosphere, also see the eprint: arXiv:0906.3110 

\noindent Hiremath, K. M., 2009b, see the arXiv eprint

\noindent Howe, R., Cristensen-Dalsgaard, J., Hill, F., {\em et. al.,} 2000, Science, 287, 2456

\noindent Ichimota, K., et. al., 1985, Nature, 316, 422

\noindent Javaraiah, J and Komm, R. W., 1999, Sol. Phys, 184, 41

\noindent Javaraiah, J., 2000, Ph. D. Thesis, Study of Sun's rotation and solar activity, Bangalore University, India 

\noindent Kane, R. P., Vats, H. O. and Sawant, H. S., 2001, Sol. Phys., 2011, 181

\noindent Kane, R. P., 2002, Sol. Phys., 205, 351

\noindent Kilic, H., 2009, Sol. Phys., 255, 155

\noindent Knaack, R.; Stenflo, J. O.; Berdyugina, S. V., 2005, A\&A, 438, 1067

\noindent Krivova, N. A. \& Solanki, S. K., 2002, A \& A, 394, 701

\noindent Lara, A; Borgazzi, A; Mendes, Odim, Jr.; Rosa, R. R.; Domingues, M. O, 2008;
Solar Physics, 248, 155
 
\noindent Lean, J. L \& Brueckner, G. E., 1989, ApJ, 337, 568

\noindent Lean, J. L., 1990, ApJ, 363, 718

\noindent Lean, J. L, 1991, Rev. Geophys., 29, 4, 505

\noindent Leighton, R. B, 1960, in Proc. IAU Symp.,  12, 321

\noindent Lou, Yu-Qing; Wang, Yu-Ming; Fan, Zuhui; Wang, Shui; Wang, Jing Xiu, 2003, MNRAS,
345, 809

\noindent Obridko, V. N and Shelting, B. D., 2007, Advan in Space Res, 40, 1006

\noindent Oliver, R., Carbonell, M. \& Ballester, L., 1992,  Sol. Phys.,  137, 141

\noindent Olive, J.P, Chaloupy, M \& Schweitzer, H., 2002, IAU coll 188, (ESA SP-505), p. 41

\noindent Pap, J., Tobiska, W. K. \& Bouwer, S. D., 1990, Sol. Phys., 129, 165

\noindent Ribes, E., Merlin, Ph.,  Ribes, J.-C and Barthalot, R., 1989, An Geophys, 7, 321	

\noindent Rieger, E. et. al., 1984, Nature, 312, 623

\noindent Scargle, J., 1989, ApJ, 343, 874

\noindent Singh, J. \& Prabhu, T. P., 1985, Sol. Phys., 97, 203

\noindent Sturrock, P. A., Walther, G  and  Wheatland, M. S., 1997, ApJ, 491, 409

\noindent Vecchio, A.; Carbone, V., 2009, A\&A, 502, 981

\noindent Verma, V. K.; Joshi, G. C.; Uddin, Wahab; Paliwal, D. C., 1991, A\&AS, 90, 83

\noindent Wolff, C. L., 1983, ApJ, 264, 667
 
\noindent Zang, J., {\em et. al.,} 2001, ApJ, 452, 2001

\clearpage

\begin{figure}[h]
\begin{center}
\psfig{figure=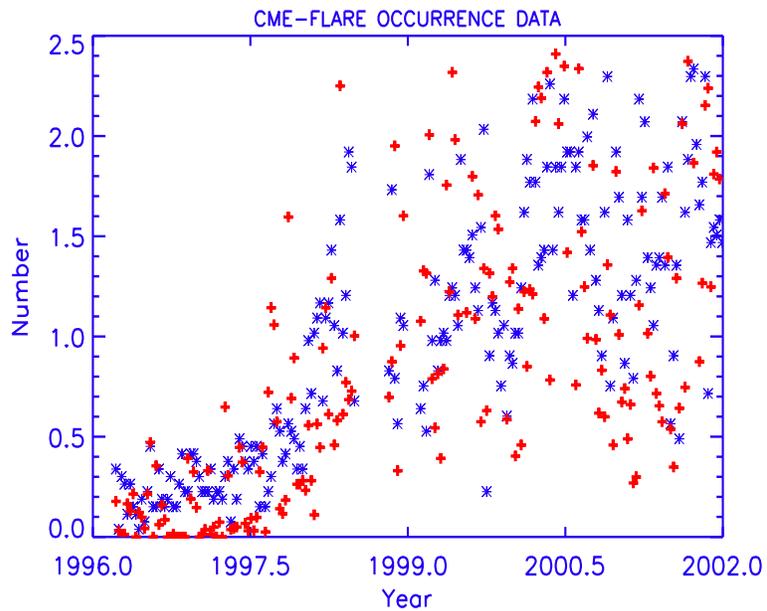,height=8.0cm,width=10.0cm}
\end{center}
\caption{This graph illustrates the number of occurrences of CME
and Halpa-flares normalized to their respective means are plotted
with respect to year.
The CME occurrences are represented with the stars in blue color and the flare
occurrences are represented as plus signs with red color.}
\end{figure}

\clearpage 

\begin{figure}[h]
\begin{center}
\psfig{figure=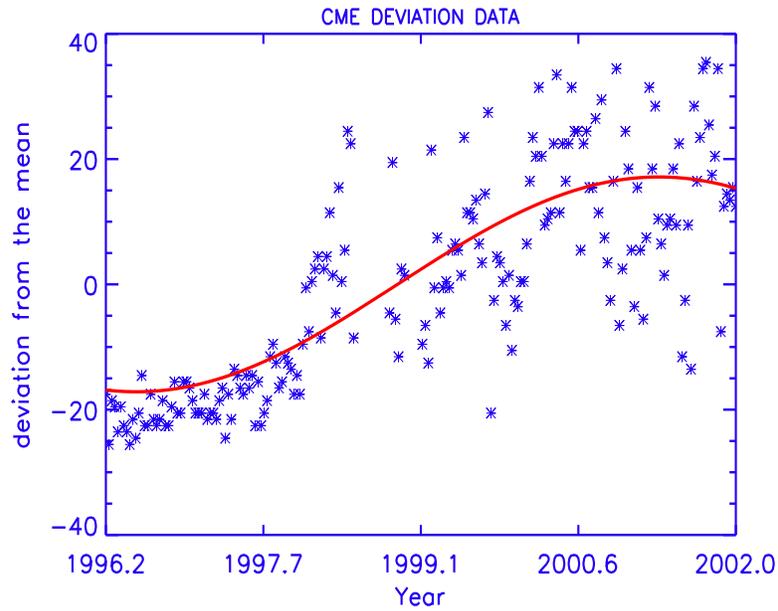,height=8.0cm,width=10.0cm}
\end{center}
\caption{The figure illustrates the deviation from the mean of the 
CME occurrence with respect to year.
The red continuous line superposed on the deviation data is fit of a sine curve
of the form $y=Asin(\omega t + \phi)$, where $y$ is the deviation of the CME occurrences
from the mean, $A=-17.03$ is amplitude, $\omega = 2\pi/T$ (T= 9.3 yrs) is
the frequency and $\phi (=1.22$ radians) is the phase. } 
\end{figure}
\clearpage

\begin{figure}[h]
\begin{center}
\psfig{figure=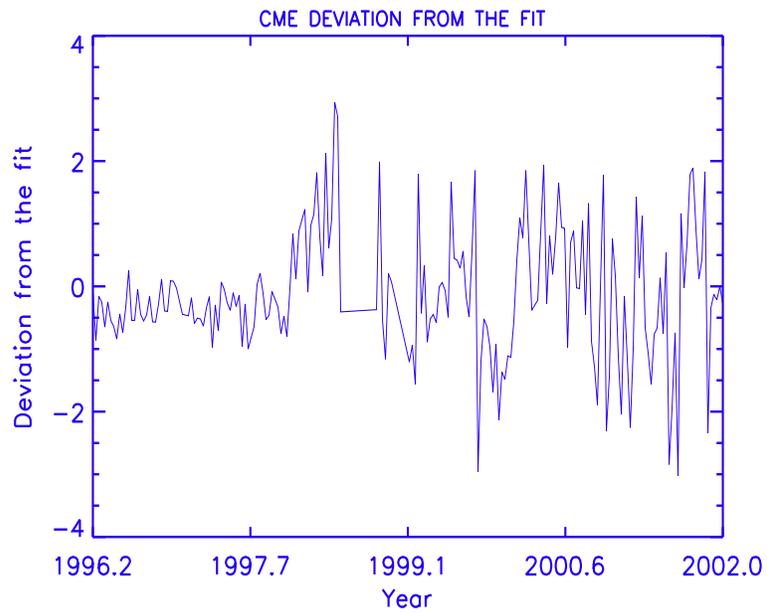,height=8.0cm,width=10.0cm}
\end{center}
\caption{The figure represents the difference from the sine curve obtained
from the least square fit and deviation from the mean of the CME occurrence
 plotted with respect to year. Notice a gap in the data during July 1998-Sept 1998
and Jan 1999.}
\end{figure}

\clearpage

\begin{figure}[h]
\begin{center}
\psfig{figure=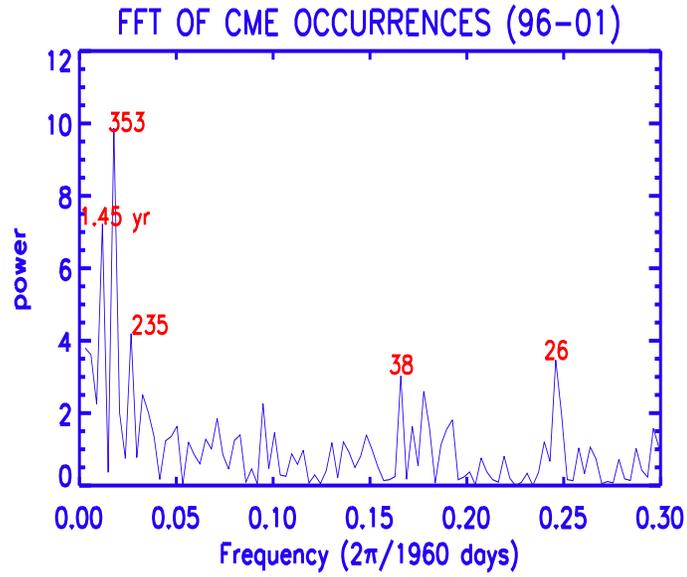,height=8.0cm,width=10.0cm}
\end{center}
\caption{The figure illustrates the Scargle Fourier
power spectrum of the CME occurrences with the gaps during July 1998-Sept
1998 and Jan 1999.  The numbers in red color are periods whose powers are
more than 3 $\sigma$ significance level. Except 1.45 yr period, rest of the
detected periods are in days. }
\end{figure}

\clearpage

\begin{figure}[h]
\begin{center}
\psfig{figure=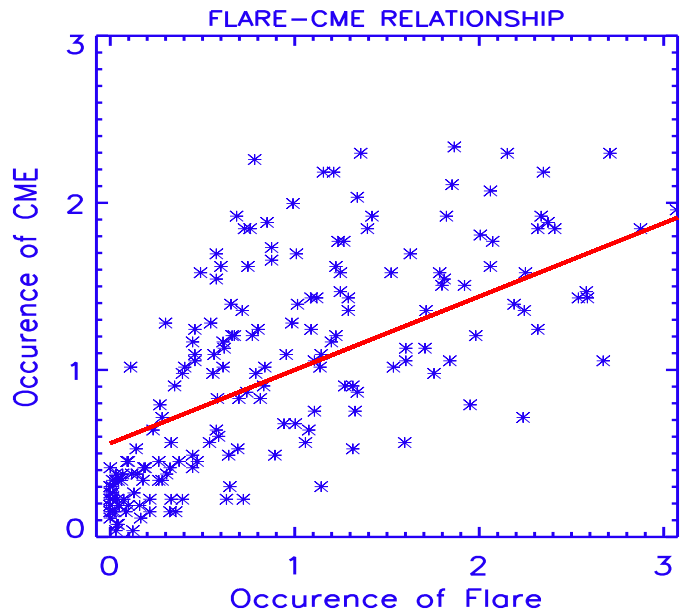,height=8.0cm,width=10.0cm}
\end{center}
\caption{The figure represents a good correlation (65 \%) between
the CME and the flare occurrences normalized to their respective means. The superposed
continuous straight line in red color is a linear least square fit for both the data
set. The least square fit yields a straight line $CME = A + B(flare)$, where $A=0.56\pm0.01$
and $B=0.44\pm0.08$. From this relation and the known flare occurrences during the months
of Jul 1998-Sept 1998 and Jan 1999, we fill the CME occurrence data during those 
gapped months. }
\end{figure}

\clearpage

\begin{figure}[h]
\begin{center}
\psfig{figure=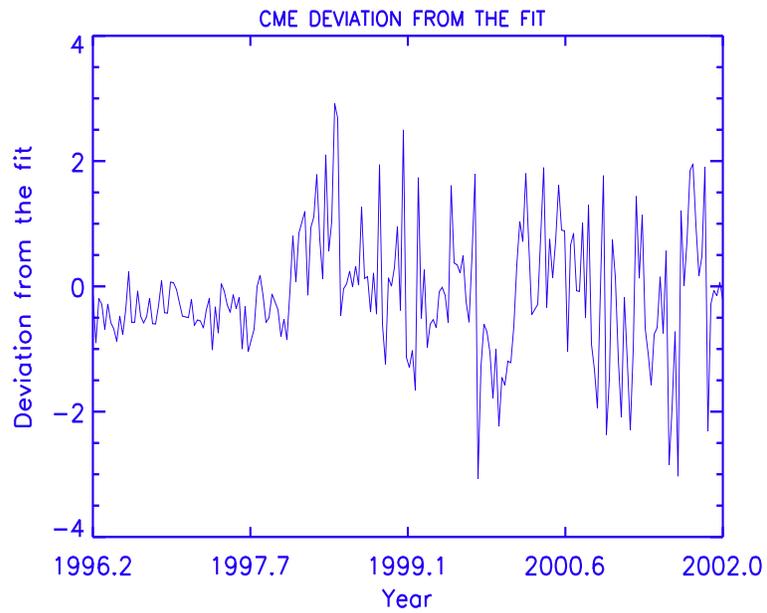,height=8.0cm,width=10.0cm}
\end{center}
\caption{This figure is similar to that of the {\bf Fig 3} with a
difference that the occurrence data is filled during the gapped months 
(July 1998-Sept 1998 and Jan 1999) of CME observations. }
\end{figure}
\clearpage

\begin{figure}[h]
\begin{center}
\psfig{figure=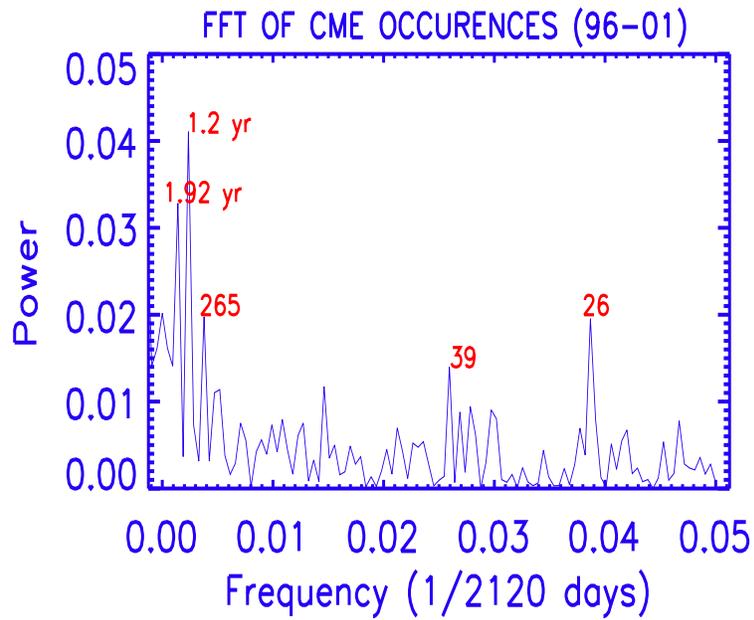,height=8.0cm,width=10.0cm}
\end{center}
\caption{This figure illustrates the power spectrum obtained
from the FFT analysis with equal intervals in time of the filled CME occurrence.
 The numbers in red colors are the
periods detected with significance of their powers more than 3 $\sigma$ level. 
The periods 1.92 and 1.2 are in years and rest of the detected periods
are in days. }
\end{figure}
\clearpage

\begin{figure}[h]
\begin{center}
\psfig{figure=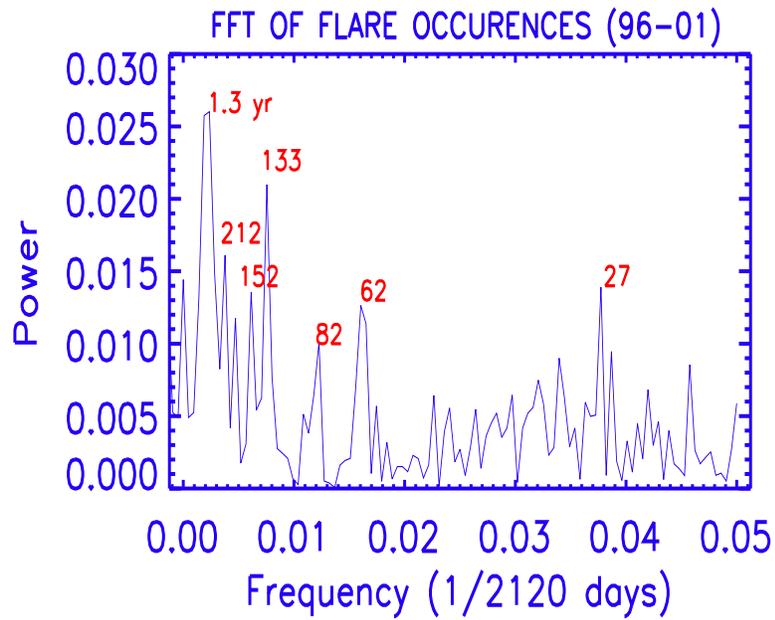,height=8.0cm,width=10.0cm}
\end{center}
\caption{This figure illustrates the power spectrum of the flare occurrences
from the FFT analysis with equal intervals in time. The numbers in red colors are the
periods detected with significance of their powers more than 3 $\sigma$ level. 
The period 1.3 is in years and rest of the detected periods are in days.}
\end{figure}
\clearpage

\begin{figure}[h]
\begin{center}
\psfig{figure=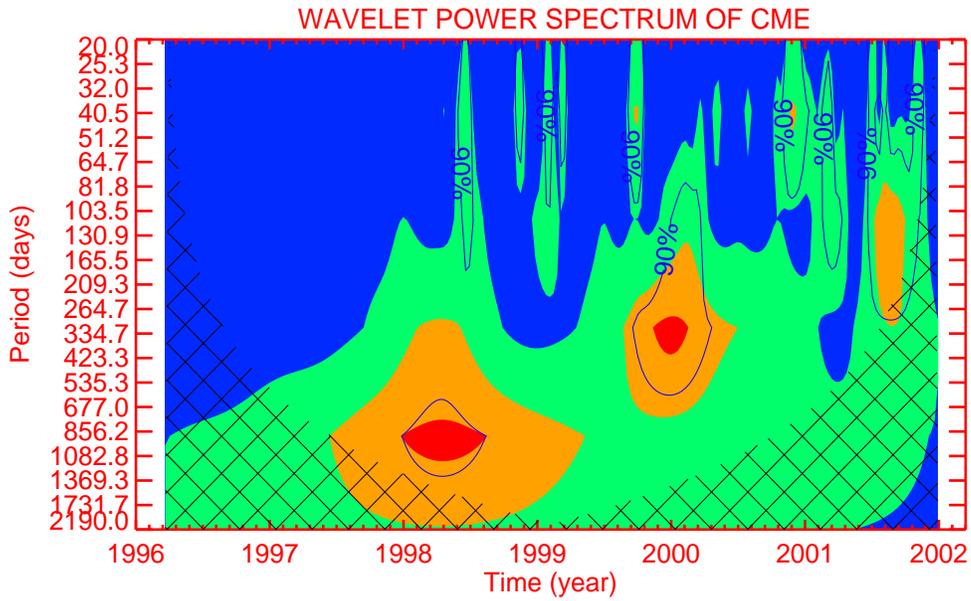,height=8.0cm,width=13.0cm}
\end{center}
\caption{The figure represents the wavelet power spectrum
of the CME occurrences with filled data set.
The regions in the interior of the contours with a light blue
color represents concentration of the power with a
confidence level of $90 \% $ significance and the power which is insignificant is
present in the crossed hatched grey regions.}
\end{figure}
\clearpage

\begin{figure}[h]
\begin{center}
\psfig{figure=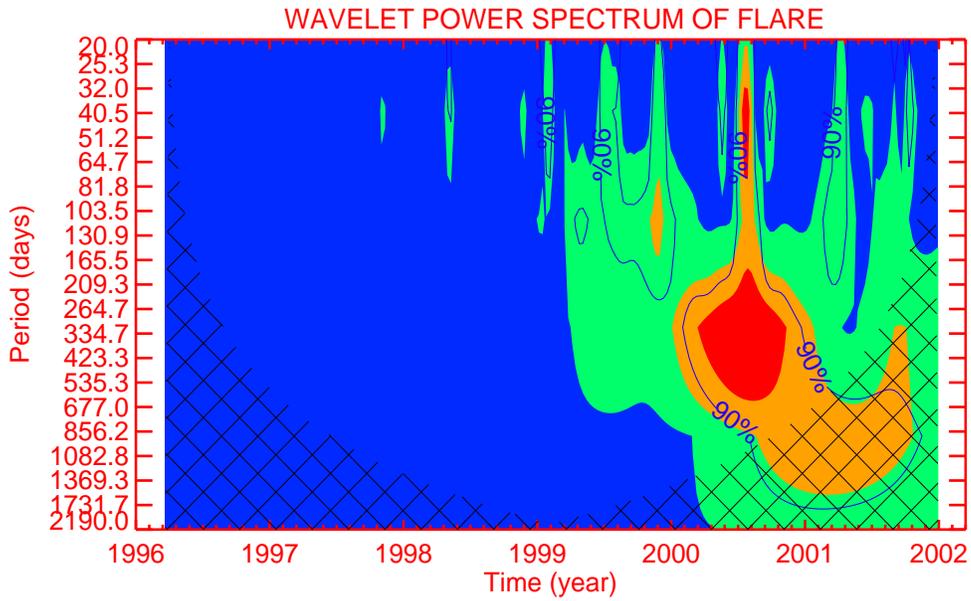,height=8.0cm,width=13.0cm}
\end{center}
\caption{This illustration represents the wavelet power spectrum
of the flare occurrences.  The regions in the interior of the contours with a light blue
color represents concentration of the power with a
confidence level of $90 \% $ significance and the power which is insignificant is
present in the crossed hatched grey regions.}
\end{figure}
\end{document}